\documentclass{jpsj-suppl}
\usepackage{txfonts} 

\newcommand{\zaa}{Astron.~Astrophys.}
\newcommand{\zapj}{Astrophys.~J.}
\newcommand{\zprd}{ Phys.~Rev. D}
\newcommand{\zprc}{ Phys.~Rev. C}
\newcommand{\zprl}{Phys.~Rev.~Lett.}
\newcommand{\zjcap}{J. Cosmol. Astropart. Phys.}
\newcommand{\zadndt}{At. Data Nucl. Data Tables}
\newcommand{\znim}{Nucl.~Inst.~and~Meth.}
\newcommand{\znpa}{Nucl.~Phys.~A}

\newcommand{\zmnras}{Mon. Not. R. Astron. Soc.}

\newcommand{\etal}{{\it et al.}}
\newcommand{\obh}{$\Omega_{\mathrm{b}}{\cdot}h^2$}

\newcommand{\deu}{${\rm D}$}
\newcommand{\tro}{$^3{\rm He}$}
\newcommand{\qua}{$^4{\rm He}$}
\newcommand{\six}{$^{6}{\rm Li}$}
\newcommand{\sep}{$^{7}{\rm Li}$}

\newcommand{\ddn}{D(d,n)$^3$He}
\newcommand{\ddp}{D(d,p)$^3$H}
\newcommand{\dpg}{D(p,$\gamma)^3$He}
\newcommand{\hag}{$^3$He($\alpha,\gamma)^7$Be}

\title{Primordial Nucleosynthesis}

\author{Alain \textsc{Coc}}

\inst{Centre de Sciences Nucl\'eaires et de Sciences de la
Mati\`ere  (CSNSM), CNRS/IN2P3, Univ.~Paris-Sud, Universit\'e~Paris--Saclay, F--91405 Orsay Campus, France}

\email{coc@csnsm.in2p3.fr}

\recdate{August 10, 2016}

\abst{Primordial or big bang nucleosynthesis (BBN)  is now a parameter free theory whose predictions are in good overall 
agreement with observations. However, the \sep\ calculated abundance is significantly higher  than the one 
deduced from  spectroscopic observations. Most solutions to this {\em lithium problem} involve
a source of extra neutrons that inevitably leads to an increase of the deuterium abundance. This seems now to be excluded by
recent deuterium observations that have drastically reduced the uncertainty on D/H and also calls for improved precision  
on thermonuclear reaction rates.
}

\kword{Big bang nucleosynthesis, Lithium, Deuterium, Thermonuclear reaction rates}

\begin{document}
\maketitle

\section{Introduction}

Primordial nucleosynthesis is one of the three historical strong evidences for the hot big bang model. Its last free parameter,
the baryon--to--photon ratio of the Universe, is now deduced from observations of the anisotropies of 
the cosmic microwave background radiation (CMB), with a precision better than one percent \cite{Planck15}.
There is a good agreement between the primordial abundances of \qua\ and \deu,  deduced from observations, and from primordial nucleosynthesis calculations. 
However, the \sep\ calculated abundance is significantly higher  than the one deduced from  spectroscopic observations. 
Solutions to this problem that have been considered include stellar surface depletion of lithium,
nuclear destruction during BBN or  solutions beyond the standard model (see \cite{Fie11} for a review). 
Experiments have now excluded a conventional
nuclear physics solution (see e.g. \cite{Ham13} and references therein), even though a few uncertain reaction rates could marginally affect Li/H predictions.   
This lithium problem  has recently worsened. Most non--conventional solutions lead to an increase of deuterium production. 
However, recent deuterium observations have drastically reduced the uncertainty on primordial D/H abundance \cite{Coo14}, excluding
such increase. 
With a precision of 1.6\% on the observed D/H value \cite{Coo14}, comparison with BBN predictions requires that the uncertainties  
on thermonuclear reaction rates governing deuterium destruction be reduced to a similar level.

\section{Recent results}

In our latest work \cite{Coc15}, we adopted for the baryon--to--photon ratio, the constraints obtained with the largest set of CMB data (TT,TE,EE+lowP), 
without any external data, giving  \obh\   = 0.02225$\pm$0.00016, together with $N_\nu$=3 for the number of neutrino families and 
$\tau_{\rm{n}}$ = 880.3$\pm$1.1~s \cite{PDG} for the neutron lifetime. The nuclear reaction rates are those listed in \cite{Coc12b}
with only a few updates listed below. 
Recently, Hou \etal\ \cite{Hou15} have re--evaluated the $^7$Be(n,$\alpha)^4$He cross section, based on $^4$He($\alpha$,n)$^7$Be, 
$^4$He($\alpha$,p)$^7$Li and $^7$Li(p,$\alpha)^4$He experimental data, using  charge symmetry and/or detailed balance principles. 
An improved evaluation of the $^3$He($\alpha,\gamma)^7$Be reaction rate  has been published \cite{deB14}, using a 
Monte--Carlo based R--matrix analysis. The effect of these re--evaluations can be seen between columns 2 and 3 in Table~\ref{t:heli},
while columns 4 shows the effect of the re--evaluated reaction rates \cite{Coc15} for deuterium destruction (see \S~\ref{s:nucl}).  
Column 5 represent the results of a Monte Carlo calculation \cite{Coc15} as described in \cite{Coc12b,Coc14} but with these few
new rates, to be compared with observations in Column 6.  
 
\begin{table*}[htbp!] 
\caption{\label{t:hlix} Primordial abundances compared to observations.}
\begin{center}
\begin{tabular}{ccccccccc}
\hline
& a & b & c & d & Observations & Cyburt et al. \cite{Cyb16} \\
\hline
$Y_p$      &0.2482&0.2482&0.2484& 0.2484$\pm$0.0002 & 0.2449$\pm$0.0040\cite{Ave15}& 0.24709$\pm$0.00025 \\
\deu/H   ($ \times10^{-5})$ &2.635&2.635&2.452& 2.45$\pm$0.05 & 2.53$\pm$0.04 \cite{Coo14}& 2.58$\pm$0.13 \\
\tro/H    ($ \times10^{-5}$)  &1.047&1.047&1.070& 1.07$\pm$0.03 & 1.1$\pm$0.2 \cite{Ban02}& 1.0039$\pm$0.0090\\
\sep/H ($\times10^{-10}$)   &5.040&5.131&5.651& 5.61$\pm$0.26 &   1.58$^{+0.35}_{-0.28}$\cite{Sbo10}& 4.68$\pm$0.67\\
\hline
\end{tabular}\\
Baseline (a), update of $^7$Be(n,$\alpha$)\qua\ \cite{Hou15} and \hag\ \cite{deB14} rates (b), together with \ddn, \ddp\ and \dpg\ new rates \cite{Coc15} (c), 
Monte Carlo (1$\sigma$) (d)
from Coc \etal\ \cite{Coc15}.
\end{center}
\label{t:heli}
\end{table*}

It is apparent in Table~\ref{t:heli} that the lithium prediction is higher than observations by a factor $\approx$3.5, but also
that deuterium predictions are only marginally compatible with recent observations \cite{Coo14}. 
The last columnn in Table~\ref{t:heli} displays the results of a very recent review by Cyburt et al. \cite{Cyb16} showing
only small differences with our work. They virtually disappear when the new rates discussed above are adopted in both 
calculations (Tsung-Han Yeh, {\em priv. comm.}), except for \qua, due, apparently, to different corrections to
the weak rates.

\section{\sep\ and \deu\ nucleosynthesis}
\label{s:nucl}

\begin{figure}[h!]
\begin{center}
\includegraphics[width=.8\textwidth]{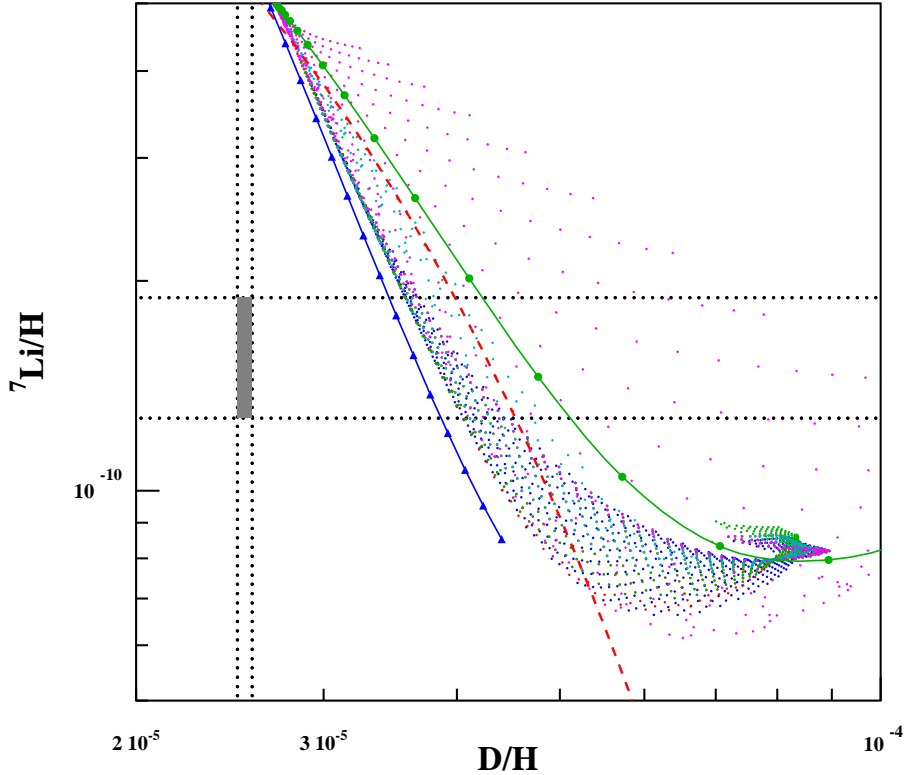}
\caption{Lithium--deuterium anti--correlation in BBN induced by different models involving neutron injection (dots: update of
Fig.~9 in Ref. \cite{Coc14a}, green circles:  Fig.~7 in Ref. \cite{Coc14a} and blue triangles, Fig.~12 in Ref. \cite{Coc12a}). 
The horizontal and vertical dotted lines  rerepresents the observational  Li/H \cite{Sbo10} and D/H \cite{Coo14} constraints  
while the dashed line is a qualitative explanation of the anti--correlation.}
\label{f:cor}
\end{center}
\end{figure}

For the CMB deduced baryon--to--photon ratio, \sep\ is produced indirectly by $^3$He($\alpha,\gamma)^7$Be,
where $^7$Be will much later decay to \sep, while $^7$Be is destroyed by  $^7$Be(n,p)$^7$Li(p,$\alpha)^4$He.
The solutions to the lithium problem generally rely on an increased late time neutron abundance 
to boost $^7$Be destruction  through the $^7$Be(n,p)$^7$Li(p,$\alpha)^4$He channel. 
Figure~\ref{f:cor}  summarizes the results on \sep\ and \deu\ predictions by different models
than include late time neutron injection aiming at reducing the $^7$Be+$^7$Li production, but at the expense 
of \deu\ overproduction. These models involve mirror neutrons, dark matter decay or annihilation  \cite{Coc14a} 
or coupled variation of constants (affecting the  $^1$H(n,$\gamma$)$^2$H rate) \cite{Coc12a} as extra neutron sources.
These extra neutrons, inevitably, also boost the D and $^3$H production through the  
$^1$H(n,$\gamma)^2$H and $^3$He(n,p)$^3$H channels, respectively \cite{Kus14}. 
The dashed curve \cite{Coc15} represent an approximation  [Eq.~7.4 in \cite{Coc15}]  of the interplay 
between Li/H and D/H when neutrons are injected towards the end of BBN ($T<$0.5~GK).
In this approximation, the flow through the $^3$He(n,p)$^3$H reaction is neglected but it shows up in the lower limit ($\approx0.6\times10^{-10}$ in Fig.~\ref{f:cor})
reached in Li/H: \sep\ is produced by the $^3$He(n,p)$^3$H($\alpha,\gamma$)$^7$Li reaction at low temperature when the 
$^7$Li(p,$\alpha)^4$He is reaction is less efficient. The figure shows that many models  \cite{Oli12,Kus14} are able to bring the lithium abundance within the
observational limits but at the expense of an increased D/H abundance ($\approx4\times10^{-5}$), now excluded by observations. 
In addition, it was noted by Kusakabe et al. \cite{Kus14} that the ratio of $^1$H+n to $^7$Be+n cross sections increases with
energy, rendering less efficient the injection of non--thermalized neutrons (from heavy relic decays e.g.  \cite{Oli12}) for destroying $^7$Be 
without overproducing deuterium.

\begin{figure}[tbh]
\begin{center}
\includegraphics[width=.8\textwidth]{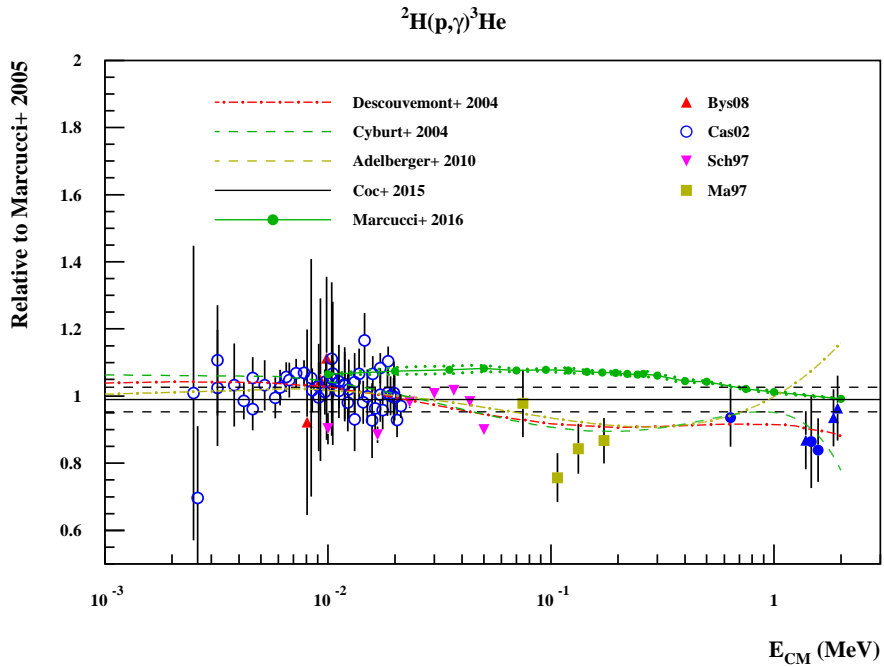}
\caption{Ratio of experimental \cite{Bys08b,Cas02,Sch97,Ma97}, fitted \cite{Des04,Cyb04,Ade11,Coc15} and new theoretical \cite{Mar16}
$S$--factors to the theoretical one \cite{Mar05}; 
the horizontal lines correspond to the theoretical $S$--factor scaled by $\alpha\pm\Delta\alpha$ \cite{Coc15}. (Systematic uncertainties 
in the range 4.5--9\% are not shown.)}
\label{f:dpg}
\end{center}
\end{figure}

Leaving aside this unsolved lithium mystery, the precision of 1.6\% (or better \cite{Coo16}) on the observed D/H value \cite{Coo14},  requires that the uncertainties  
on the \dpg, \ddn\ and \ddp\ rates that govern deuterium BBN destruction be reduced to a similar level. Indeed, a +1\% variation of these rates 
induces a respectively -0.32, -0.46 and -0.54\% variation of D/H \cite{Coc15}.
Achieving such a precision on nuclear cross sections is a very difficult task: data from different experiments need to be combined keeping in mind the 
importance of systematic uncertainties, in particular concerning the absolute normalization. Two main philosophies are found in evaluations of reaction rates:
$i$) follow closely experimental $S$--factor experimental data or $ii$) use theoretical input for the shape of the $S$--factor. We chose the second option,
with Marcucci \etal\ \cite{Mar05} [\dpg] and Arai \etal\ \cite{Ara11} [\ddn\ and \ddp] as theoretical $S$--factors, keeping the normalization  ($\alpha$) as a free 
parameter that has to be determined by comparison with experimental data. The procedure we followed \cite{Coc15} for the \dpg\ 
reaction was $i$) to select experimental datasets \cite{Bys08b,Cas02,Sch97,Ma97} for which systematic uncertainties were provided, $ii$) determine
for each data set,  by $\chi^2$ minimization,  the normalization factor to be applied to the theoretical $S$--factor of  Marcucci \etal\ \cite{Mar05}, 
$iii$) add quadratically the systematic uncertainties and $iv$) perform a weighted average of the normalization factor.          
We obtained $\alpha=0.9900\pm0.0368$ for this factor, that was subsequently used to scale the  Marcucci \etal\  $S$--factor, and calculate
the thermonuclear \dpg\ reaction rate and associated uncertainty. This result is quite robust given the data and the theoretical $S$--factor, as verified using 
bayesian techniques instead \cite{bayes}. 
Comparison between experimental data, fits and theories is displayed in Fig.~\ref{f:dpg} 
(normalized to the  Marcucci \etal\ (2005) theoretical $S$--factor). It shows that previous fits \cite{Ade11,Des04,Cyb04} were driven down 
by the scarce data at BBN energies. This is not the case anymore when the theoretical energy dependence of Marcucci \etal\ (2005) is assumed. 
However, an improved calculation of the 
$S$--factor by Marcucci \etal\ (2016) \cite{Mar16} lies significantly above the previous calculation: if one applies the same renormalisation 
 method one finds $\alpha=0.915\pm0.038$.   
We used the same procedure for \ddn\ and \dpg\ except that the theoretical $S$--factor is taken from Arai \etal\ \cite{Ara11}.
All three reaction rates are higher than previous evaluations  at BBN temperatures leading to a decrease in the D/H prediction, as shown
 in Table~\ref{t:heli}. In addition, if we now use the theoretical $S$--factor from Marcucci \etal\ (2016), we obtain an additional reduction of
 $\Delta$(D/H) = -0.072$\times10^{-5}$ that vanished if we rescale it ($\alpha$=0.915) to fit experimental data.

\section{Summary and conclusions}

As conclusions, we list below our comments regarding frequently asked questions concerning big bang nucleosynthesis.

\begin{itemize}
\item {\em There is no nuclear solution to the lithium problem.} Extensive sensitivity studies \cite{Coc12b} have not
identified reactions, beyond those already known, that could have a strong impact on lithium nucleosynthesis.
Unknown resonances that could sufficiently increase the cross sections of reactions that destroy $^7$Be were
not found experimentally (see e.g. \cite{Ham13} and references therein) and in any case would have too low strengths \cite{Bro12} because of the Coulomb barrier.      
\item However, {\em without solving the lithium problem}, uncertainties affecting a few reaction rates, like \hag, may still affect the lithium production. 
The role of the $^7$Be(n,$\alpha)^4$He reaction is presently assumed to be negligible with respect to  
$^7$Be(n,p)$^7$Li. However, depending on the results of ongoing experiments (these proceedings), it could reduce the
lithium production by a few percents. The up to now overlooked $^7$Be(n,p$\gamma$)$^7$Li channel could also 
have a similar effect \cite{Hay16}.
\item {\em The effect of electron screening or modification of decay lifetime is negligible.} For reactions of interest to BBN, screening 
affects the laboratory cross sections at too low energies [e.g. $\lesssim$ 20~keV for \ddp\ or  $^3$He(d,p)$^4$He] to affect
measurement at BBN energies [$\approx$100~keV], on the one hand.  On the other hand, the effect screening during BBN is completely negligible 
\cite{Wan11,Fam16}. 
It is well known that the lifetime of $^7$Be that decays by electron capture is modified in a plasma \cite{Sim13}. However, because of 
the Boltzmann suppression factor, at $T<$0.5 GK, the electron density is too low to provide the required reduction factor of $\approx$3000 on 
its 53 days half-life.  
\item Many exotic solutions to the lithium problem have been investigated (e.g. \cite{Yam14}), but most rely on extra neutron sources 
that {\em overproduce deuterium to levels
now excluded by observations} \cite{Coo14,Coo16}. Few solutions beyond the Standard Model that do not suffer from this drawback
are left, e.g. \cite{Gou16}.     
\item {\em Stellar physics solutions requires a uniform reduction of surface lithium over a wide range of effective temperature and metallicity.}
With some fine--tuning, this could be achieved by the combined effects of atomic diffusion and turbulence in the outer layers of
these stars \cite{Ric05}, or by lithium destruction, followed by a self--regulated re-enrichment of lithium by late time  accretion \cite{Fu16}.    
\item {\em There is no \six\ problem anymore.} 
A few years ago, observations \cite{Asp06}  of \six\ in a few metal poor stars had suggested the presence of a plateau, at 
typically  \six/H $\approx10^{-11}$, orders of magnitude higher than the BBN predictions of 
 \six/H $\approx1.3\times10^{-14}$ \cite{Ham10}.
The uncertainties on the D$(\alpha,\gamma)^6$Li cross section have been experimentally constraint  by a LUNA measurement \cite{And14}
and by theory \cite{Muk16} confirming the BBN value. However,
later, the observational \six\  plateau has been questioned  
due to line asymmetries which were neglected in previous abundance analyses.
Hence, there is no remaining evidence for a plateau at very low metallicity \cite{Lin13} that can be used to derive a primordial \six\  abundance.
\item With the high precision on D/H observations, the \dpg, \ddn\ and \ddp\ rates need to be known at the percent level! 
This demands accurate measurement at BBN energies where data are scarce (see Fig.~\ref{f:dpg}), to be compared with theories.
The theoretical work of Arai \etal\ \cite{Ara11} was focused on low energies and does not correctly reproduce the \ddn\ and \ddp\  
experimental data above $\approx$600~keV. It is highly desirable that these calculations be extended up to $\approx$2~MeV,
to cover the range of experimental data and BBN energies. 
\end{itemize}

\section*{Acknowledgments}
I am indebted to my collaborators on these topics:  Pierre Descouvemont, 
St\'ephane Goriely,  Fa\"{\i}rouz Hammache, Christian Iliadis, Keith Olive, Patrick Petitjean, Maxim Pospelov, Jean-Philippe Uzan, and especially to Elisabeth Vangioni.



\begin{thebibliography}{99}

\bibitem{Planck15} 
{\it{Planck  }}  Collaboration XIII, P.A.R. Ade, N. Aghanim,  \etal,
{\tt arXiv:1502.01589 [astro-ph.CO]}.

\bibitem{Fie11} B.D. Fields, 
Annu. Rev. Nucl. Part. Sc. {\bf 6}  (2011) 47.

\bibitem{Ham13} F. Hammache, A. Coc, N. de S\'er\'eville, I. Stefan, P. Roussel, \etal,
\zprc\  {\bf 88} (2013) 062802(R).

\bibitem{Coo14} R. Cooke, M. Pettini, R.A. Jorgenson, M.T. Murphy, and C.C. Steidel, 
\zapj\  {\bf 781} (2014) 31.

\bibitem{Coc15} A. Coc,  P. Petitjean, J.-Ph.~Uzan, E.~Vangioni, P. Descouvemont, C. Iliadis and R. Longland,
\zprd\ {\bf 92} (2015) 123526.

\bibitem{PDG} 
K.A. Olive \etal\ (Particle Data Group), 
Chin. Phys. C {\bf 38} (2014)  090001 {\tt URL: http://pdg.lbl.gov}.

\bibitem{Coc12b} A. Coc, S. Goriely, Y. Xu, M. Saimpert, and E. Vangioni,
\zapj\ {\bf 744} (2012) 158.

\bibitem{deB14} R. J. deBoer, J. G\"orres, K. Smith, \etal,
\zprc\ {\bf 90}  (2014) 035804.

\bibitem{Coc14} A. Coc,   J.-Ph.~Uzan and E.~Vangioni,
\zjcap\ {\bf 10} (2014) 050.

\bibitem{Cyb16} R.H. Cyburt, B.D. Fields, K.A. Olive, and T.-H. Yeh, Rev. Mod. Phys.  {\bf 88} (2016) 015004.

\bibitem{Ave15} E. Aver, K.A. Olive, and  E.D. Skillman, \zjcap\ {\bf 07} (2015) 011.

\bibitem{Ban02} T. Bania, R. Rood, and D. Balser, 
Nature {\bf 415}  (2002) 54.

\bibitem{Sbo10} L. Sbordone, P. Bonifacio, E. Caffau \etal,
 \zaa\ {\bf 522} (2010) A26.

\bibitem{Hou15} S.Q. Hou, J.J. He, S. Kubono, and Y.S. Chen,
\zprc\ {\bf 91} (2015) 055802.

\bibitem{Coc14a} A. Coc,  M. Pospelov, J.--P.~Uzan, and E.~Vangioni,
\zprd\ {\bf 90}  (2014) 085018.

\bibitem{Coc12a}  A. Coc, P. Descouvemont, K. Olive, J.--P. Uzan, and E. Vangioni, 
\zprd\ {\bf 86}  (2012) 043529. 

\bibitem{Kus14} M. Kusakabe, M.--K. Cheoun, and K. S. Kim,
\zprd\ {\bf 90} (2014) 045009.

\bibitem{Oli12}  K.A. Olive, P. Petitjean, E. Vangioni, and J. Silk, 
 \zmnras\ {\bf 426}  (2012) 1427.

\bibitem{Coo16}R.J. Cooke, M. Pettini, K.M. Nollett and R. Jorgenson,  {\tt arXiv:1607.03900 [astro-ph.CO]},
{\em to appear} in \zapj.

\bibitem{Mar05} 
L.E. Marcucci, M. Viviani, R. Schiavilla, A. Kievsky, and S. Rosati,
\zprc\ {\bf 72}  (2005) 014001. 

\bibitem{Ara11} K. Arai, S. Aoyama, Y. Suzuki, P. Descouvemont, and D. Baye,
\zprl\  {\bf 107}  (2011) 132502.

\bibitem{Bys08b} V.M. Bystritsky, V.V. Gerasimov, A.R. Krylov,  \etal,
\znim\ A {\bf 595}  (2008)  543.

\bibitem{Cas02} C. Casella, H. Costantini, A. Lemut, B. Limata, R. Bonetti, \etal,
\znpa\ {\bf 706} (2002) 203.

\bibitem{Ma97} L. Ma, H.J. Karwowski, C.R. Brune, Z. Ayer, T.C. Black, \etal,
\zprc\ {\bf\ 55}  (1997) 588.

\bibitem{Sch97}  G.J. Schmid, B.J. Rice, R.M. Chasteler, \etal, 
\zprc\ {\bf 56}  (1997) 2565.

\bibitem{bayes} C. Iliadis, K. S. Anderson, A. Coc, F. X. Timmes and S. Starrfield, {\tt arXiv:1502.01589 [astro-ph.SR]},
{\em to appear} in \zapj.

\bibitem{Des04} P. Descouvemont, A. Adahchour, C. Angulo, A. Coc, and E. Vangioni-Flam, 
\zadndt\ {\bf 88} (2004) 203  and {\tt http://pntpm.ulb.ac.be/bigbang/}.

\bibitem{Ade11} E. G. Adelberger, A. Garc\'{\i}a, R. G. Hamish Robertson, \etal,
Rev. Mod. Phys. {\bf 83}   (2011) 195.

\bibitem{Cyb04} R.H. Cyburt, 
\zprd\ {\bf 70}  (2004) 023505.

\bibitem{Mar16} L. E. Marcucci, G. Mangano, A. Kievsky, and M. Viviani,
\zprl\ {\bf 116}  (2016) 102501.

\bibitem{Bro12} C. Broggini, L. Canton, G. Fiorentini and F.L. Villante,
\zjcap\ {\bf 06} (2012) 030.

\bibitem{Hay16} S. Hayakawa \etal, 
Proceedings of the {\em Sicily-East Asia Workshop on Low Energy Nuclear Physics},
Il  Nuovo Cimento, {\em to be published} 

\bibitem{Wan11} B. Wang, C. A. Bertulani, and A. B. Balantekin,
\zprc\ {\bf 83}  (2011) 018801.

\bibitem{Fam16} M. A. Famiano, A. B. Balantekin, and T. Kajino,
\zprc\ {\bf 93} (2016) 045804. 

\bibitem{Sim13} S. Simonucci, S. Taioli, M. Busso and S. Palmerini,
\zapj\ {\bf 764} (2013) 118.

\bibitem{Yam14} D.G. Yamazaki, M. Kusakabe, T. Kajino, G.J. Mathews, and M.--K. Cheoun,
\zprd\ {\bf 90} (2014) 023001. 

\bibitem{Gou16}A. Goudelis, M. Pospelov, and J. Pradler,
\zprl\ {\bf 116}  (2016)  211303.

\bibitem{Ric05}
O. Richard, G. Michaud and J. Richer, \zapj\ {\bf 619} (2005) 538. 

\bibitem{Fu16}
X. Fu, A. Bressan, P. Molaro and P. Marigo,
\zmnras\ {\bf 452} (2015) 3256. 

\bibitem{Asp06} M. Asplund, D.L. Lambert, P.E. Nissen \etal, \zapj\ {\bf 644} (2006) 229. 

\bibitem{Ham10}
F. Hammache, M. Heil, S. Typel, D. Galaviz, K. S\"ummerer et al.,  
\zprc\  {\bf 82}  (2010) 065803. 

\bibitem{And14} M. Anders, D. Trezzi, R. Menegazzo, M. Aliotta, A. Bellini et al.,
\zprl\ {\bf  113}  (2014) 042501.

\bibitem{Muk16} A. M. Mukhamedzhanov, Shubhchintak and C. A. Bertulani,
\zprc\ {\bf 93}  (2016) 045805.

\bibitem{Lin13} 
K. Lind, J. Melendez, M. Asplund et al., \zaa\ {\bf554} (2013) 96.





\end{thebibliography}
\end{document}